\documentclass[
    numbers,
    sort&compress
]{elsarticle}

\usepackage[utf8]{inputenc}
\usepackage[T1]{fontenc}
\usepackage[english]{babel}
\PassOptionsToPackage{hyphens}{url}
\usepackage{hyperref}
\usepackage{amsmath}
\usepackage{csquotes}
\usepackage[inline]{enumitem}
\usepackage{graphicx}
\usepackage{wrapfig}
\usepackage[font+=footnotesize]{subcaption}
\usepackage[
    update,
    outdir=./gfx/converted/,
    suffix=
]{epstopdf}
\usepackage{booktabs}
\usepackage[capitalize]{cleveref}
\Crefname{lstlisting}{Listing}{Listings}

\usepackage[nolist]{acronym}
\usepackage{xspace}
\usepackage{xcolor}

\usepackage{listings}
\usepackage{tikz}
\usepackage{pgf-umlsd}
\usepgflibrary{arrows}

\usepackage{siunitx}

\usepackage[style=iso]{datetime2}
\usepackage{placeins}

\newcommand{\ie}{i.e.,\xspace}
\newcommand{\eg}{e.g.,\xspace}
\newcommand{\etal}{et~al.\xspace}
\newcommand{\dy}{Dolev-Yao\xspace}
\newcommand{\did}{defence in depth\xspace}

\let\oldfootnote\footnote
\def\footnote{\ifhmode\unskip\fi\oldfootnote}

\newcommand{\signal}[1]{Signal~#1}
\newcommand{\point}[1]{Point~#1}
\newcommand{\track}[1]{Track~#1}

\newif\ifcomments
\definecolor{seceng}{HTML}{B90F22}
\definecolor{markus}{HTML}{008000}

\newcommand{\mycomment}[3]{    
    \ifcomments
        {%
            \noindent%
            \ifhmode%
                \unskip%
            \fi%
            \color{#1}%
            \textbf{\scriptsize{#2:}} #3%
        }%
    \fi
}

\usepackage[breakable,most]{tcolorbox}
\tcbset{
    breakable,
    title=Notes,
    enhanced
}
\usepackage{comment}
\ifcomments
    \newenvironment{mybox}{\begin{tcolorbox}}{\end{tcolorbox}}
\else
    \excludecomment{mybox}
\fi

\journal{Critical Infrastructure Protection}

\hypersetup{
    pdftitle = {Rule-based Anomaly Detection for Railway Signalling Networks},
    pdfauthor = {Markus Heinrich, Arwed Gölz, Tolga Arul, Stefan Katzenbeisser},
    pdfkeywords = {safety and security co-engineering, railway signalling, rule-based anomaly detection, cyber-physical system, semantic attack, critical infrastructure protection}
}

\begin{document}

\begin{frontmatter}

    \title{Rule-based Anomaly Detection for Railway Signalling Networks}

    \author[1]{Markus Heinrich\corref{cor1}}
    \ead{heinrich@seceng.informatik.tu-darmstadt.de}
    \author[1]{Arwed Gölz}
    \author[1]{Tolga Arul}
    \ead{arul@seceng.informatik.tu-darmstadt.de}
    \author[2]{Stefan Katzenbeisser}
    \ead{stefan.katzenbeisser@uni-passau.de}

    \address[1]{Technische Universität Darmstadt, Department of Computer Science, Germany}
    \address[2]{University of Passau, Faculty of Computer Science and Mathematics, Germany}

    % !TEX root = ../rule-based-anomaly-detection.tex
% !TEX spellcheck = en_GB
% !TEX encoding = UTF-8 Unicode

\begin{abstract}
We propose a rule-based anomaly detection system for railway signalling that mitigates attacks by a \dy attacker who is able to inject control commands and to perform semantic attacks.
The system as well mitigates the effects of a compromised signal box that an attacker uses to issue licit but mistimed control messages.
We consider an attacker that could cause train derailments and collisions, if our countermeasure is not employed.
We apply safety principles of railway operation to a distributed anomaly detection system that inspects incoming commands on the signals and points.
The proposed anomaly detection system detects all attacks of our model without producing false positives, while it requires only a small amount of overhead in terms of network communication and latency compared to normal train operation.
\end{abstract}


\end{frontmatter}

% ---------- Section Inputs ----------
% !TEX root = ../rule-based-anomaly-detection.tex
% !TEX spellcheck = en_GB
% !TEX encoding = UTF-8 Unicode

\begin{acronym}
    \acro{ci}[CI]{critical infrastructure}
    \acro{cots}[COTS]{commercial off-the-shelf}
    \acro{cps}[CPS]{cyber-physical system}
    \acro{fe}[FE]{field element}
    \acro{ics}[ICS]{industrial control system}
    \acro{ids}[IDS]{intrusion detection system}
    \acro{mac}[MAC]{message authentication code}
    \acro{nids}[NIDS]{network intrusion detection system}
    \acro{rasta}[RaSTA]{Rail Safe Transport Application}
    \acro{tds}[TDS]{train detection system}
\end{acronym}

% !TEX root = ../rule-based-anomaly-detection.tex
% !TEX spellcheck = en_GB
% !TEX encoding = UTF-8 Unicode

\section{Introduction}
\label{sec:introduction}

% Motivation
% Problem statement COTS
Railway signalling systems, like many other \acp{ics}, increasingly leverage \ac{cots} hardware, public networks and open protocols in order to exploit the benefits of standardized products.
At the same time, this transformation exposes railway signalling to a range of new security threats~\cite{schlehuber2017challenges,schlehuber2017security,bloomfield2012secure,temple2017train,valdivia2018cybersecurity} because standard products lower the threshold for attacks by vulnerabilities becoming public more likely and being easier to exploit.
Physical protection of railway signalling infrastructure is virtually impossible due to its spatial extension along the railway tracks.
This provides a large attack surface for the attacker with little chance of being caught and sufficient time to perform attacks on signalling devices and networks.

% Safety/Security co-engineering
Securing safety-critical signalling infrastructure is a difficult task because security measures must not interfere negatively with the safety functionality in order that railway transportation remains safe and keeps the safety certification.
Interference could, for example, result from a security application introducing delay in the safety's communication, such that safety is unable to meet response time requirements and subsequently violates the system's fail-safe property.
Another cause of friction between railway safety and security is the difference in the duration of the system lifecycle.
In safety, a system is subject to a long lasting certification process and is assumed safe forever after completing it.
In security on the contrary, vulnerabilities and attacks become public permanently such that systems are required to be updated and patched frequently.
This is incompatible with today's safety certification because the system cannot be left vulnerable for the time a safety re-certification lasts.

% Solution: Defence in Depth
A solution to bring safety and security together is to wrap the safety functionality into a protective layer of security~\cite{schlehuber2017challenges} without the need to modify safety systems, safety applications, or safety protocols.
The discussed difficulty to integrate security in safety, the large number of devices, its spatial extension, and the lack of physical protection constitute the complexity of the \ac{ci} railway signalling.
This also increases the chance that employed security measures fail or are compromised by an attacker.
A single countermeasure cannot cover the full complexity of the attack surface and the increasing number of attacks on \acp{ci}.
A \did concept with multiple different defence strategies mitigates the effect and makes railway signalling more resilient to attacks.
In this way, a security measure can be ineffective, failing, or compromised while the overall system remains protected.

% One building block: anomaly detection
We investigate anomaly detection on the controllers of railway signalling as one building block among many of a \did concept.
Other security measures should cover communication integrity and authentication, hardware integrity against physical tampering with devices, a separation kernel to isolate different pieces of software, and health monitoring during runtime as discussed by Heinrich \etal~\cite{heinrich2019security}.

% Contribution
We study sophisticated semantic attacks that make use of licit control commands to set railway signals and points to a state that may cause train accidents.
Firewalls and input validation on the receiver cannot mitigate an effect of these commands that -- on their own -- are licit but arrive mistimed at the controller.
Authenticating command and control network communication does not suffice in railway signalling because authentication is difficult to introduce in safety protocols, could be compromised by the attacker or might be ineffective if the attacker controls the signal box that is authorized to dispatch commands.

We show that it is possible to enhance the IT security resilience of railway signalling networks against application layer attacks misusing licit control commands by a rule-based anomaly detection system.
Safe train operation is ensured by the signal box of a railway station that controls all mutable infrastructure elements and prohibits any hazardous configuration and unsafe train movement.
However, the safety system does neither ensure the integrity of the signal box nor the authentic transmission and execution of the signal box's commands to the \aclp{fe}.
Therefore, licit commands can be manipulated, delayed, dropped, resent, reordered, or injected.

To mitigate this threat, we use the safety logic encoded in the signal box as template for a distributed, rule-based anomaly detection system.
The system validates critical commands on the individual receiving \acl{fe} against the state of relevant, adjacent \aclp{fe} that are neighbours in the physical track topology.
Furthermore, we describe attack patterns of a strong attacker on the railway system who possesses expert knowledge of railway signalling and has moderate resources available to perform attacks.
Moreover, we discuss the important aspect of safety and security co-engineering: how to deal with detected security incidents while maintaining the fail-safety and freedom of interference of the system.
Security measures should be closely tied to safety functionality to maintain operation of the safety system in the presence of disruptions~\cite{young2014integrated}.
Finally, we implement our anomaly detection system and evaluate it on a dataset of authentic railway command and control traffic.

\begin{mybox}
Solving the conflict of objectives between safety and security has been investigated with formal methods~\cite{troubitsyna2016towards,feuser2014dependability}
Message authentication for railway-specific transport protocols has been investigated to provide an initial level of protection~\cite{pepin2016risk,heinrich2018security}.
Railway signalling, with its safety criticality and specific security requirements, demands a \did concept to ensure safe public transportation~\cite{heinrich2019security}.

Arweds Thesis:~\cite{goelz2018rule-based}

\paragraph{Contribution}

\begin{itemize}
    \item Signal Mapping? needs more evaluation
\end{itemize}

\end{mybox}

% !TEX root = ../rule-based-anomaly-detection.tex
% !TEX spellcheck = en_GB
% !TEX encoding = UTF-8 Unicode

\section{Related Work}
\label{sec:related_work}

We are the first to explore the field of anomaly detection in railway signalling as there are only few publications examining cybersecurity in railway at all.
There are several publications on other \acp{cps} and \acp{ics} with comparable properties that are relevant to our investigation of defending against semantic attacks.

% own publications
Schlehuber \etal~\cite{schlehuber2017security,schlehuber2017challenges} have described the challenges of protecting safety-critical railway signalling from cyberattacks.
Railway signalling systems are designed and certified for multiple decades of lifetime assuming they do not experience significant modification.
Frequent updates of security measures (\eg vulnerability patches, malware signatures) would violate the certification if not designed properly.
To avoid re-certification in case of a change in the security measures, Schlehuber \etal propose a \enquote{shell concept} comprising a protective \emph{security shell} around the safety functionality.
With the shell concept, security and safety functionality is separated such that security can be updated independently from safety, and security does not interfere negatively with the safety functionality.
Heinrich \etal~\cite{heinrich2019security} propose a hardware platform according to the shell concept that allows the execution of safety and security functionality on the same hardware without interference by the help of a separation kernel.
The rule-based anomaly detection proposed in this article can be one of the security measures deployed on this platform that interacts with the safety functionality which controls the railway signalling \aclp{fe}.

% valdivia2018cybersecurity
Valdivia \etal~\cite{valdivia2018cybersecurity} highlight that safety in railways is well studied and has produced several standards for safety-critical systems.
However, similar to Schlehuber \etal, they identify that railway cybersecurity has not experienced sufficient attention yet.
Similar to safety, standards are required to harmonize railway cybersecurity protection.
They as well describe that, due to its duration, re-certification of the safety system must be avoided if a security update is applied.
As a solution to this update problem, they describe a \ac{nids} architecture that physically separates safety and \ac{nids} (security).
The \ac{nids} does not require safety certification because it operates on a communication channel comprising safety communication with built-in interference detection.
We propose an anomaly detection that works on the safety communication as well, but on the same hardware platform as the safety application, as proposed by Heinrich \etal~\cite{heinrich2019security}.
The hardware platform avoids the physical gap between safety system and security system which again could be the surface to mount an attack.
Heinrich \etal have shown that virtual separation is sufficient to operate safety-critical and non-safety-critical applications.

% fovino2011critical
Fovino \etal~\cite{fovino2011critical} show that current firewall generations cannot detect attacks that are based on licit commands.
In particular, the firewalls fail to determine whether a licit command arrives mistimed and transforms the system in a critical state.
Moreover, a series of licit commands could be chained together to move the system to a critical state.
The authors propose to enrich the firewall with information about the architecture and the state of the protected system.
We use equivalent information for railway signalling, but use it to enrich the actuators themselves and not a distinguished firewall.

% caselli2015sequence
Similar to Fovino \etal, Caselli \etal~\cite{caselli2015sequence} describe an \ac{ids} that defends against sequence attacks.
They utilize a Markov chain model of the controlled \ac{cps} as countermeasure.
Their controlled system is a water treatment and purification facility that uses Modbus communication.
Sequence attacks are a specific type of semantic attack that concern the misplacement of system operation events that are licit if considered individually.

% jin2006anomaly
Jin \etal~\cite{jin2006anomaly} employ the semantics of the controlled process to infer anomaly detection rules similar to the idea behind our work.
Their work is based on electricity networks, where they developed invariants to verify across defined locations of the network.
The Linear Invariant Checker and Bus-zero-sum Checker are derived from physical laws that are always valid in an electrical grid.

% temple2017train
Temple \etal~\cite{temple2017train} study the impact of an attacker on train service.
The attacker targets availability and integrity of balises that are used to determine the exact position of a train for precise stopping.
They consider train operation as a \ac{cps} because a train is governed by physical laws while influenced by control actions.
Their countermeasure includes a train braking model of the physical world to mitigate the influence of missing or forged balise location information.

\begin{mybox}

safety/security, railway, intrusion detection:~\cite{valdivia2018cybersecurity}

Argument of~\cite{carcano2011multidimensional}: industrial systems are subject of safety analysis processes, therefore the possible critical states are well documented.
Attackers goal is forcing a transition of the system into a critical state.

\end{mybox}

% !TEX root = ../rule-based-anomaly-detection.tex
% !TEX spellcheck = en_GB
% !TEX encoding = UTF-8 Unicode

\section{Model}
\label{sec:model}

We first outline a typical signalling network infrastructure that we consider in this article.
Then, we explain the capabilities of an attacker in this infrastructure by devising an appropriate attacker model.

\subsection{Infrastructure Model}
\label{sec:infrastructure_model}

In railway signalling, a signal box communicates with \acp{fe} via an Ethernet/IP network within the area of a railway station.
In our railway infrastructure model, we consider three types of \acp{fe}: light signals, points and \acp{tds}.
The latter monitor the vacancy of track sections.
To ensure safe train movement through the infrastructure, the signal box receives reports on the vacancy of track sections from the \acp{tds} and sends according commands to the points and light signals.
In turn, the points and light signals report the execution of a command with their updated state back to the signal box.
We call this type of communication the \textbf{safety channel} because it happens for the safety of train operation.
The signal box keeps track of the states of all \acp{fe} in its supervised area.
It updates the states accordingly after commanding a state change and receiving the respective confirmation.
A human traffic controller supervises train operation by requesting the signal box to set the infrastructure to allow the desired train movement.
The signal box validates any requested state change against its internal states.
If the state change would create an unsafe state (potential derailment or collision), the request is rejected.
In the following section, we discuss why this validation for safety reasons does not suffice from a security perspective.

\begin{figure}[htb]
    \centering
    \includegraphics[width=\textwidth]{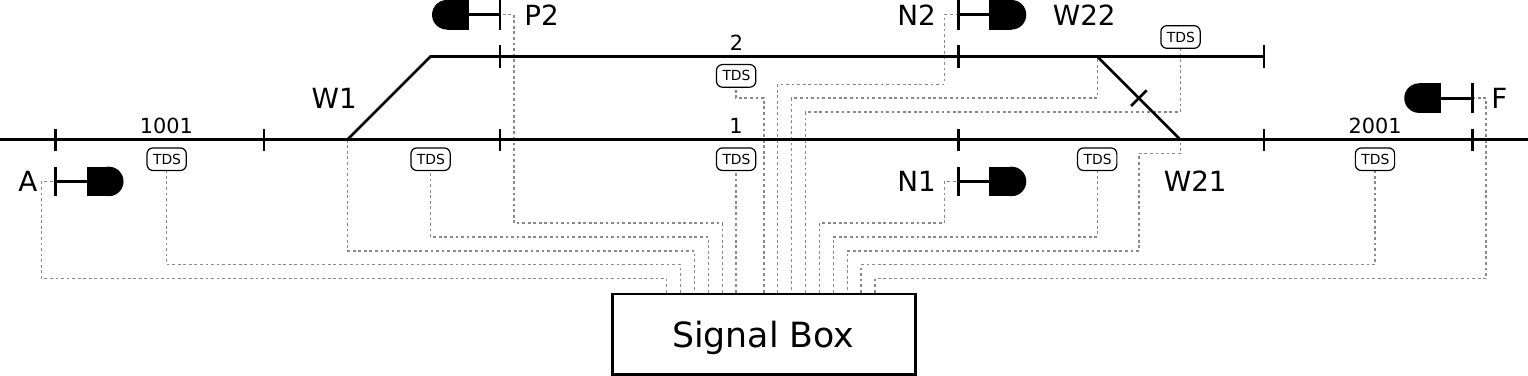}
    \caption[Example Topology.]{Example topology of a railway signalling network.
        Black lines depict the railway track.
        Thin dashed lines depict safety channels.
        A, Px, Nx, F depict signals.
        Wx depict points.
        TDS indicates \aclp{tds}.
    }
    \label{fig:signalling_network}
\end{figure}

An example topology of a railway signalling network including all types of considered \acp{fe} is shown in \cref{fig:signalling_network}.
We use it to showcase our anomaly detection in this article.
Thin dashed lines indicate the safety channels between signal box and \acp{fe}.
In general, one signal box operates all the \acp{fe} in the area of a single railway station.

\subsection{Attacker Model}
\label{sec:attacker_model}

% Introduction
Current signalling networks transmit commands and reports between the signal box and \acp{fe} via closed but otherwise unprotected networks.
On top of the Ethernet, IP and UDP stack, a transport protocol called \ac{rasta} is standardised in Germany, that provides only limited security properties to protect against cyberattacks~\cite{heinrich2018security}.
Authentication and integrity have already been identified as security requirements of railway signalling and several solutions have been proposed to mitigate security risks~\cite{schlehuber2017security,schlehuber2017challenges,heinrich2019security}.

% Reasons for security
However, for several reasons it is important to deploy further security measures in signalling networks.
First, following a \did concept~\cite{apta2013securing}, it is good practice to design systems with multiple layers of security, in case parts of the defence concept (\eg message authentication) are broken or circumvented by an attacker.
Second, due to the certification processes required by safety authorities, it might be impossible to introduce security measures in the safety channel, leaving the communication unprotected.
Railway safety is a strictly regulated domain where any change of the system undergoes a certification process whose duration is incompatible with the frequent changes IT security requires to maintain adequate protection.
Current practice is to put a protective \emph{security shell}~\cite{schlehuber2017challenges} around the safety system (the signal box and \ac{fe} control), leaving the safety unchanged during a security patch to avoid the need for re-certification.
In contrast, if the attacker successfully penetrates the \emph{security shell}, the safety system is fully exposed and unprotected.
This underlines the need for a \did concept which the presented anomaly detection is a part of.
From a safety perspective, we consider derailments and collisions as the attacker-provoked hazards that must be mitigated by the \emph{security shell}.
This view is supported in other literature, such as Bloomfield \etal~\cite{bloomfield2016risk}.
In general, the approach of the attacker is to gain control over a \ac{fe} or to desynchronise the internal state of the signal box from reality such that a hazardous command is issued by the signal box due to wrong information about a \ac{fe}'s state.

% DY attacker
In this article, we consider a \dy attacker~\cite{dolev1983security} who gained access to the signalling network but cannot break cryptographic primitives such as encryption.
For the safety channel, this implies that the attacker can monitor any message and therefore infer the state of every \ac{fe} and the location of every train in the supervised area.
As messages, we consider control commands and reports that adhere to the application and transmission layer protocols used in railway signalling.
These message are: command and report to set a signal aspect, command and report to switch a point, and the report of a \ac{tds} whether a track section is occupied.
Furthermore, without a functional \emph{security shell}, the attacker can suppress any message between the signal box and a \ac{fe} as well as inject arbitrary messages that will be accepted by the receiving entity.
We consider this feasible because tools to spoof and inject messages in standard protocols such as Ethernet, IP and UDP are easily available, as are the definitions of railway application protocols.
Exploiting her capabilities, the attacker can provoke a derailment by switching a point while a train is running over it.
Furthermore, she could set a signal to clear with the intention of leading a train into a section that is already occupied by another vehicle, provoking a collision.
For both conditions to occur, a single command posing as the signal box to a \ac{fe} is sufficient.
Consequently, a sophisticated attacker can stay unnoticeable to any detection mechanism until she sends her first harmful message.
This requires a strong defence mechanism that is capable of detecting single harmful messages in the context of railway signalling.

% Physical attacks
We do not consider physical attacks on signalling equipment because they are easily possible due to the spatial extension of the railway network.
It is even simpler to perform physical attacks that do not qualify as cyberattacks.
Blocking the tracks is executable with less resources and expert knowledge than, for example, physically compromising the controller of a \ac{fe}.
However, physical attacks can not be scaled to a multitude of \acp{fe} compared to the attacks we describe, which can be executed on multiple targets at the same time.
Additionally, physical attacks often leave witnesses, and most importantly, an identifiable offender, whereas cyberattacks may be implemented as a thousand papercuts without any clear idea of the attacker, or, if implemented particularly well, that there even is an attacker (\eg Stuxnet).

% Signal Box Compromisator
We also consider a second attacker type in this work.
This attacker compromises the signal box of the railway station instead of attacking on network level like the \dy attacker.
With this power over the signal box, she can configure a set of \acp{fe} to provoke derailments or collisions as well.
The signal box is the single authority to send commands to the \acp{fe} via the safety channel.
Thus, message authentication on the safety channel provides no protection in this case.
This attacker type could be internal (malicious employee), a railway targeted malware (like Stuxnet) that made its way to the signal box, or a remote hacker who gained access to the signal box.

% !TEX root = ../rule-based-anomaly-detection.tex
% !TEX spellcheck = en_GB
% !TEX encoding = UTF-8 Unicode

\section{Anomaly Detection}
\label{sec:anomaly_detection}

% introduction
We explain each component of our anomaly detection system in detail and illustrate the functionality in an example where a route for a train is set.
Then, we discuss how a security alert can be treated in a safety-critical system by mapping security incidents to safety failure modes.

\subsection{Security Channel}
\label{sec:security_channel}
\begin{figure}[htb]
    \centering
    \includegraphics[scale=.7]{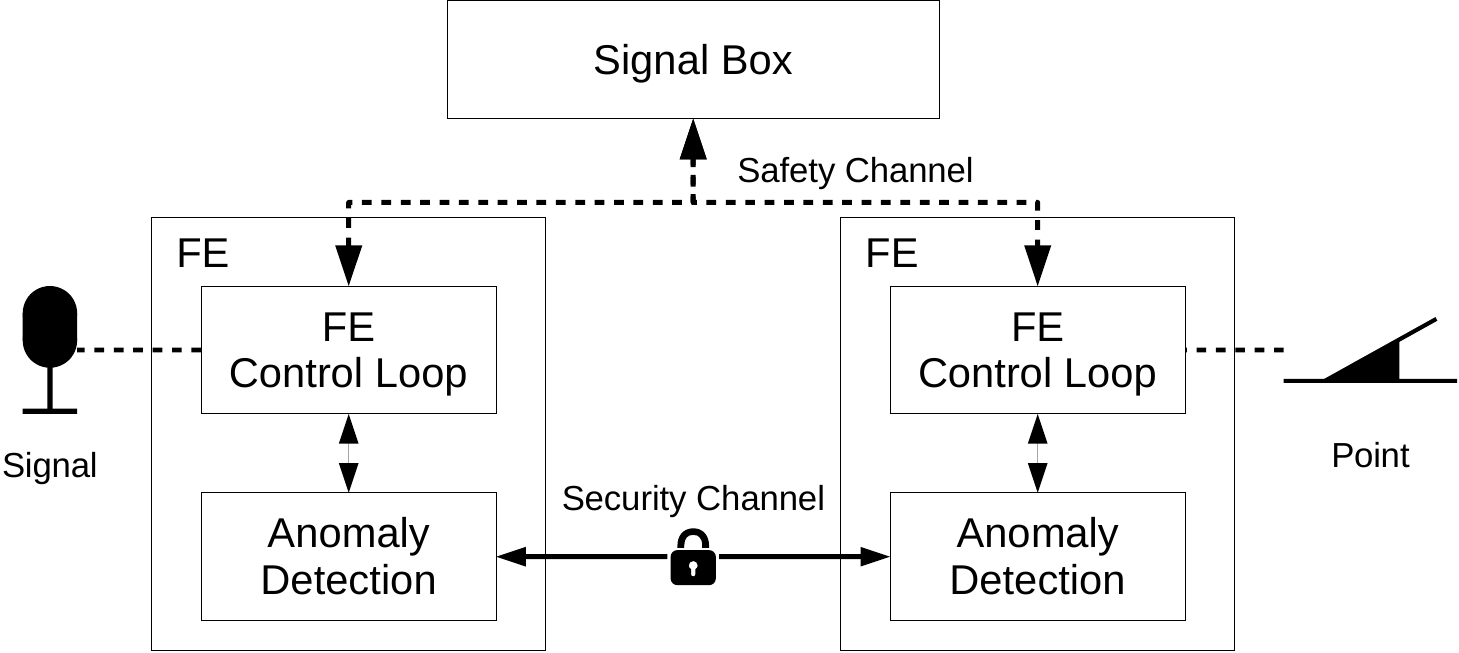}
    \caption{Overview of Signal Box, \ac{fe} control loop and anomaly detection.}
    \label{fig:component_overview}
\end{figure}

We observe that the \acp{fe} slavishly obey any command that is sent to their control loop by the signal box and cannot distinguish hazardous commands.
Our strategy to counter the shortcomings is to enhance the \acp{fe} with state information of nearby \acp{fe}.
We extend the star-shaped communication pattern of the safety channels shown in \cref{fig:signalling_network}, that stipulates signal box to \ac{fe} communication only.
To exchange additional information about the state of neighbouring \acp{fe}, we allow communication between \acp{fe}.
With this extra information, \acp{fe} can collaboratively validate a received command against the configuration of the surrounding infrastructure.
We design our anomaly detection as a process that runs in an isolated partition on the same hardware that is controlling the \acp{fe} as proposed by Heinrich \etal~\cite{heinrich2019security}.
The anomaly detection receives the messages of the safety channel and influences the control loop running in a second partition in case of a detected anomaly.
As the exchanged information and the subsequent analysis by the \ac{fe} is a security measure, we call the communication between \acp{fe} \textbf{security channel}.
\Cref{fig:component_overview} shows the control loop of two \acp{fe} interacting with the signal box via a safety channel and their anomaly detection processes interacting via a security channel.
Messages sent over the security channel need to be authenticated, such that an attacker can not alter the exchanged information.

\begin{figure}[htb]
    \centering
    \begin{subfigure}[b]{.5\linewidth}
        \centering
        \includegraphics{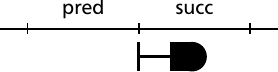}
        \caption{Signal}
        \label{fig:signal}
    \end{subfigure}%
    \begin{subfigure}[b]{.5\linewidth}
        \centering
        \includegraphics{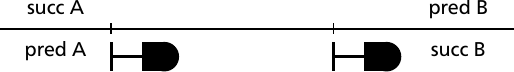}
        \caption{TrackSectionTDS}
        \label{fig:track_section}
    \end{subfigure}

    \vspace{2em}
    \begin{subfigure}[b]{\linewidth}
        \centering
        \includegraphics{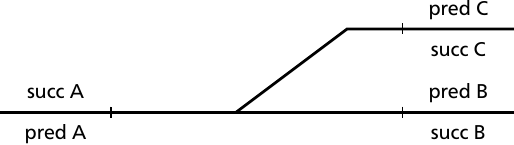}
        \caption{Point, PointTDS}
        \label{fig:point}
    \end{subfigure}

    \caption{Visualization of \aclp{fe}.}
    \label{fig:visualization_of_fe}
\end{figure}

% types of field elements
We have identified three types of \acp{fe} that we consider for anomaly detection in this paper: signals, points, and \acp{tds}.
For \acp{tds}, we distinguish two subtypes that either monitor a track section (TrackSectionTDS) or a point (PointTDS).
\Cref{fig:visualization_of_fe} gives an overview of the three types.

\begin{figure}[htb]
    \centering
    \includegraphics[width=\textwidth]{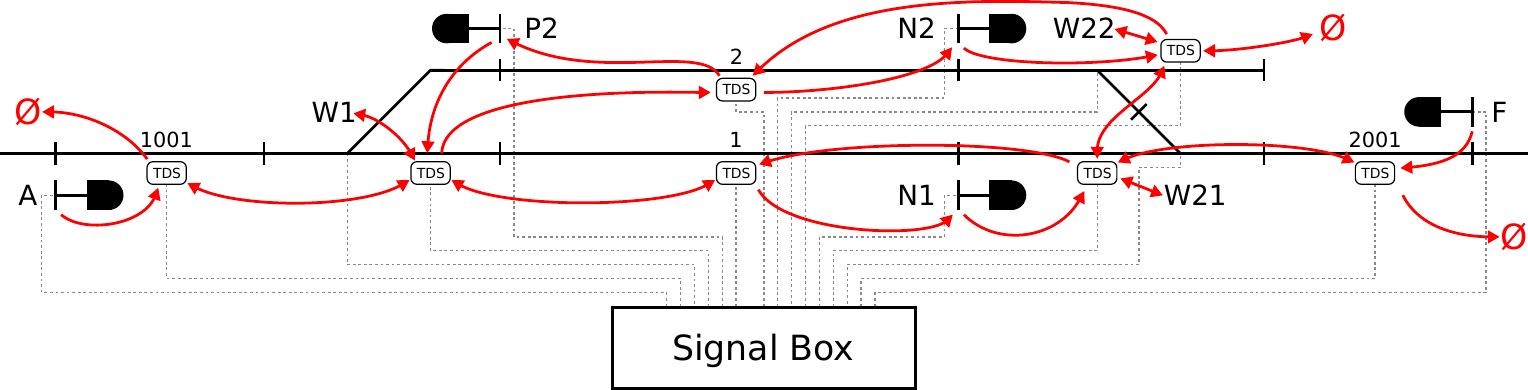}
    \caption{Topology with arrows visualizing Security Channels.}
    \label{fig:security_channel}
\end{figure}

% additional communication links / security channel
We explain the communication links (security channel) that are established by our anomaly detection.
Our analysis yields that a security channel needs to be established between immediate neighbours in the track topology and only between them.
A signal is linked to the \ac{tds} monitoring the adjacent track section (\enquote{succ} in \cref{fig:signal}).
A track section (\cref{fig:track_section}) connects tracks in two directions which we call A and B.
In both directions, it can be bounded by an incoming signal (\enquote{pred A} in \cref{fig:track_section}), an outgoing signal (\enquote{succ B}) or no particular \ac{fe} (\enquote{pred B}, \enquote{succ A}).
The section boundaries are marked with a orthogonal line to the track in \cref{fig:visualization_of_fe}.
Each section is monitored by a \ac{tds} for vacancy.
If a signal is present at the border of a track section, the \ac{tds} is linked to the signal.
Otherwise it is linked to the neighbouring \ac{tds}.
If no neighbour exists (\eg a dead-end track), no channel in that direction is established to implicitly model the end of the supervised area.
A \ac{tds} monitoring a point (\cref{fig:point}) adds a third direction, called C, corresponding to the concept we explained for track sections.
A train can only run safely over a point in relation A-C if the point is set to left, and in relation A-B if the point is set to right.
Train movement in relation B-C is impossible.
The \ac{tds} monitoring the point and the point itself require a security channel between each other as well.
All security channels of our example topology (see \cref{fig:signalling_network}) are visualized as arrows in \cref{fig:security_channel}.
Arrows show the successor relation in the direction they point.
Following an arrow backwards, represents the predecessor relation.
For example, \track{1001} is the successor of \signal{A}, while \signal{A} is also the predecessor of \track{1001}.
Arrows pointing to and from symbol~\O\xspace are an empty reference because the supervised area is left or the track is a dead-end.
Arrows point in both directions if the relations are also valid reversed.

\subsection{Detection Triggers}
\label{sec:triggers}
% triggers
Our rule-based anomaly detection utilizes two triggers to start its detection algorithm.
We have selected the triggers by analysing which commands can be exploited by the attacker to run an attack forcing the railway system into an unsafe state and thus are critical:
\begin{description}
    \item[Point Trigger:] Every time a point receives the command to change position, the point runs the anomaly detection algorithm explained in \cref{sec:algorithm_point}.
    \item[Signal Trigger:] Every time a signal receives a command to set to an aspect that permits train movement (set up a route), the signal starts the anomaly detection algorithm for a route explained in \cref{sec:algorithm_signal}.
\end{description}
We distinguish only two signal aspects in our anomaly detection: \emph{stop} and \emph{clear}.
There is a variety of signal aspects that allow a train to pass a signal.
Examples include the usual clear aspect, approach, substitution signal, drive on sight, and more.
We map all aspects that allow a train to pass the signal to the clear aspect to reduce the complexity of our anomaly detection rules.
In all cases where a train passes a signal, our anomaly detection rules are invoked.

\subsection{Detection Algorithms}
\label{sec:algorithms}

We distinguish two detection algorithms triggered by the point and the signal trigger presented in \cref{sec:triggers}, respectively.
An instance of the algorithm is executed on the controller of each \ac{fe} between entry and exit signal of a route.
Each \ac{fe} communicates with the respective instance on a neighbouring \ac{fe} via the security channel.
\ac{fe} state, locks and anomaly detection information are stored on each \ac{fe}'s controller as well.

\subsubsection{Algorithm for a Point}
\label{sec:algorithm_point}
% Behaviour of a point
The general idea behind our anomaly detection algorithm is the following:
on every command to change position, the point communicates with its associated \ac{tds} and requests the track's vacancy state.
Only if the \ac{tds} is not occupied by a train and not reserved for a route (locked), the \ac{tds} will acknowledge the request.
After receiving the \ac{tds}'s acknowledgement, the point will execute the command and switch position.

\subsubsection{Algorithm for a Route}
\label{sec:algorithm_signal}
% Behaviour of a signal
Every time a signal receives a command to set an aspect that allows train movement (\eg clear), a recursive query to the \acp{fe} along the pre-set route is started by the signal via the security channel.
The signal queries its successor \ac{tds} with a request whether it is safe to allow train movement.
If the track section is neither occupied nor reserved for another train (locked), the query is forwarded to the succeeding \ac{tds}, that in turn checks its status and forwards the query.
When the query reaches a signal (end of route) or a \ac{tds} with no successive \ac{tds} (dead-end track, yard limits reached), this \ac{fe} notifies the signal which started the query of the availability of the route.
If at one of the \acp{tds} along the path one or more conditions are violated, the query is returned immediately, informing the signal that the command should be rejected.
A successful query is illustrated and explained in \cref{sec:illustration}.

% TDS in general
On receiving a query, a \ac{tds} sets its internal state to locked to mark that it has been reserved for a route.
Once locked, a \ac{tds} will reject further availability queries, thus preventing multiple routes from overlapping on that track section and avoiding a collision.
The locked state is left when the \ac{tds} detects a train occupying its track section or the signal is set to an aspect prohibiting train movement (stop) without a train having passed it (\eg route cancelling by the traffic controller).
In this case, the signal notifies the unnecessarily locked \acp{fe} to release their locks again.
The algorithm performed by a \ac{tds} is sketched in \cref{lst:TrackSectionTDS.isAvailable}.
Calls to the method \texttt{isAvailable} are used to symbolize a query to an adjacent \ac{fe} via the security channel.

\begin{mybox}
maybe remove the code listings or find a better way to illustrate the functionality of the algorithm
\end{mybox}

\begin{lstlisting}[
    caption={TrackSectionTDS.isAvailable(), compare \cref{fig:track_section}.},
    label={lst:TrackSectionTDS.isAvailable},
    float=tbp
]
isAvailable(src):
    if state == TDSState.CLEAR:
        if locked:
            return False
        lock()
        if src == pred_a:
            if succ_b is not NULL:
                return succ_b.isAvailable(this)
            else:
                return True
        if src == pred_b:
            if succ_a is not NULL:
                return succ_a.isAvailable(this)
            else:
                return True
    return False
\end{lstlisting}

% PointTDS
A \ac{tds} monitoring a point additionally performs the lookup of the correct neighbour according to the state of the point and forwards the query accordingly, if such a neighbour exists.
The algorithm is sketched in \cref{lst:PointTDS.isAvailable}.

\begin{lstlisting}[
    caption={PointTDS.isAvailable(), compare \cref{fig:point}.},
    label={lst:PointTDS.isAvailable},
    float=tbp
]
isAvailable(src):
    if state == TDSState.CLEAR:
        if locked:
            return False
        lock()
        if src == pred_a:
            if point.state == PointState.RIGHT:
                if succ_b is not NULL:
                    return succ_b.isAvailable(this)
                else:
                    return True
            else:
                if succ_c is not NULL:
                    return succ_c.isAvailable(this)
                else:
                    return True
        if src == pred_b and point.state == PointState.RIGHT:
            if succ_a is not NULL:
                return succ_a.isAvailable(this)
            else:
                return True
        if src == pred_c and point.state == PointState.LEFT:
            if succ_a is not NULL:
                return succ_a.isAvailable(this)
            else:
                return True
    return False
\end{lstlisting}

\subsubsection{Unlocking a Route after a Reject}
% unlock on reject
If a query is rejected by one of the \acp{fe}, the \acp{fe} earlier in the query have locked themselves and need to be unlocked in order to be available for other routes.
Thus, the notification about the rejection travels back through the \acp{fe}, such that each \ac{fe} leaves the lock state.
The unlocking is necessary to keep the railway station operable.
Otherwise, more and more \acp{fe} would get locked over time, thus starving train operation.

\subsubsection{Cancelling a Route}
% cancel a route
The traffic controller can cancel a route before a train has used it, if necessary.
This case needs to be reflected by our anomaly detection because the \acp{tds} must be unlocked again.
Otherwise, the anomaly detection would raise false positives due to locked track sections that should actually be available.
This could eventually starve train operation because the infrastructure becomes unavailable for train movement.
For cancelling a route, the entrance signal of the route needs to be set to stop.
Then, the signal will send a message to the \acp{fe} of the route, notifying them to unlock.
This message must only be sent if no train has passed the signal yet, which the signal can determine by communicating with the adjacent \ac{tds}.
If the \ac{tds} has not been occupied while the signal was set to clear, a train cannot have passed the signal.
Thus, it is necessary to send the unlock notification.

\subsection{Illustration of Detection Rules}
\label{sec:illustration}
\begin{figure}[htb]
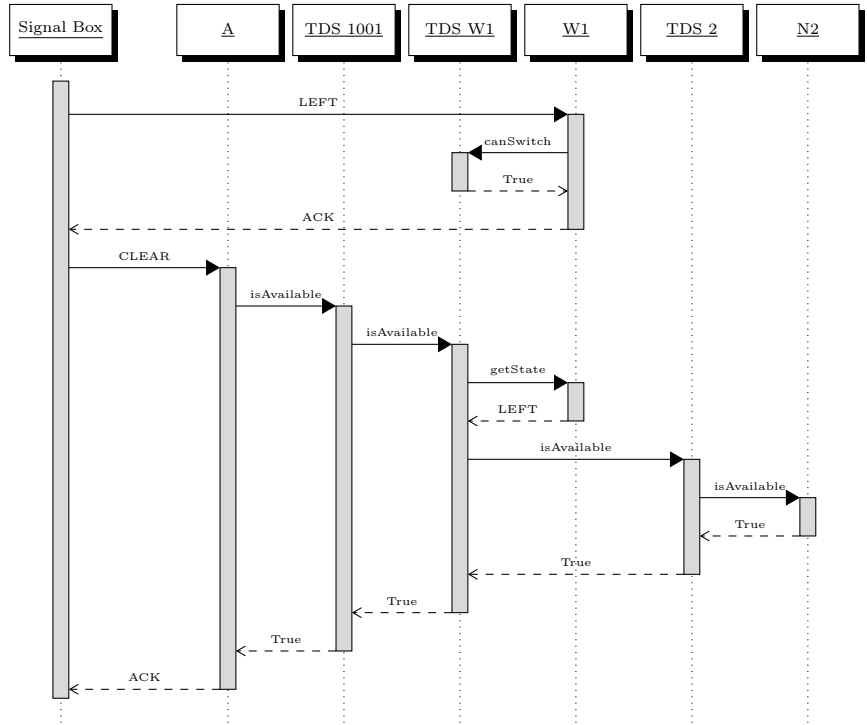

    \centering
    \tikzset{every picture/.append style={transform shape,scale=.85}}
    \begin{sequencediagram}
    \tikzset{every node/.append style={font=\scriptsize}}
        \newthread{sb}{Signal Box}
        \newinst[1]{A}{A}
        \newinst{TDS1001}{TDS~1001}
        \newinst{TDSW1}{TDS W1}
        \newinst{W1}{W1}
        \newinst{TDS2}{TDS 2}
        \newinst{N2}{N2}

        \begin{call}{sb}{\tiny LEFT}{W1}{\tiny ACK}
            \begin{call}{W1}{\tiny canSwitch}{TDSW1}{\tiny True}
            \end{call}
        \end{call}
        \begin{call}{sb}{\tiny CLEAR}{A}{\tiny ACK}
            \begin{call}{A}{\tiny isAvailable}{TDS1001}{\tiny True}
                \begin{call}{TDS1001}{\tiny isAvailable}{TDSW1}{\tiny True}
                    \begin{call}{TDSW1}{\tiny getState}{W1}{\tiny LEFT}
                    \end{call}
                    \begin{call}{TDSW1}{\tiny isAvailable}{TDS2}{\tiny True}
                        \begin{call}{TDS2}{\tiny isAvailable}{N2}{\tiny True}
                        \end{call}
                    \end{call}
                \end{call}
            \end{call}
        \end{call}
    \end{sequencediagram}

    \caption[Example Execution.]{Example execution of rule-based anomaly detection.
        Commands LEFT and CLEAR of the signal box both trigger the anomaly detection in \point{W1} and \signal{A} respectively.
    }
    \label{fig:sequence_diagram}
\end{figure}

% sequence diagram
A sequence diagram showing the safety channel and security channel communication between signal box and the \acp{fe} to set a route is depicted in \cref{fig:sequence_diagram}.
We use the infrastructure depicted in \cref{fig:signalling_network,fig:security_channel}.
The diagram assumes an approaching train from the left that starts at \signal{A}, should turn left at \point{W1}, run along \track{2} and finally stop in front of \signal{N2}.
First, the signal box commands \point{W1} to switch to left (safety channel).
\point{W1} executes the detection algorithm (security channel, \texttt{canSwitch}) and subsequently acknowledges the new state to the signal box.
Then, the signal box commands \signal{A} to show clear (safety channel).
Subsequently, \signal{A} starts the anomaly detection query (security channel, \texttt{isAvailable}) that follows the intended route until it is acknowledged by \signal{N2}.
During the query, the \ac{tds} of \point{W1} requests the state of the point before forwarding the query.
The diagram shows that the query follows the path the train will eventually take: start at \signal{A}, pass \track{1001}, \point{W1}, \track{2}, and stop before \signal{N2}.

\subsection{Handling of Alerts}
\label{sec:handling_alerts}
% dealing with alerts
Railway transportation is a \acl{ci} that is strictly regulated by safety authorities.
The criticality of the infrastructure demands careful design of any security measure to avoid interference with the safety functionality and maintain operation~\cite{young2014integrated,valdivia2018cybersecurity}.
Once the anomaly detection raises an alert, a strategy with minimal interference is required to handle it without degrading safety.
We propose that a \ac{fe} ignores a safety channel control command and keeps its state unchanged if an alert by the anomaly detection is triggered.
This maintains the fail-safe property of railway signalling because either a point is not switched or a signal remains showing a stop aspect.
This is viable because the safety system must be able to deal with \acp{fe} not reacting, as this is a failure mode considered in the safety case already.
The safety system cannot distinguish whether a message is deliberately ignored or dropped by an anomaly detection system or simply lost due to a network failure.
Additionally, our anomaly detection system is specially crafted so that it only interferes (drop or prolong execution) in commands that are not safety-critical.
However, in any case the alert can be collected and be presented to a security operator for further manual intervention.

% !TEX root = ../rule-based-anomaly-detection.tex
% !TEX spellcheck = en_GB
% !TEX encoding = UTF-8 Unicode

\section{Evaluation}
\label{sec:evaluation}

We present the dataset used for evaluation, explain our evaluation concept and discuss the results.

\subsection{Dataset}
To evaluate our rule-based anomaly detection, we use a dataset gathered over a period of three years in the signalling lab of TU Darmstadt\footnote{\url{http://www.eisenbahnbetriebsfeld.de/}}.
The signalling lab simulates real-world railway operation with tracks, signals, points, and trains in scale 1:87 and is used to qualify traffic controllers of a German railway operator.
Also, signal boxes, \acp{fe}, and their operation are accurately simulated, including the network traffic described as safety channel in \cref{sec:infrastructure_model}.
Hence, our dataset contains realistic datagrams with communication of commands and reports between the signal boxes and the \acp{fe}.
We select three railway stations from the dataset and run the recorded datagrams on an implementation of our rule-based anomaly detection.
Each station contains one signal box and is independent from the other stations in railway operation and the execution of our anomaly detection.
The stations differ in topology (layout of \acp{fe}), size (number of \acp{fe}), and complexity (types of \acp{fe} being neighbours).
Details about the datasets are provided in \cref{tab:communiation_overhead}.

\subsection{Attack Scenarios}
To validate the functionality of our anomaly detection, we define attack scenarios against the safety channel as test cases.
With the evaluation, we verify the detection capabilities of our anomaly detection model defined in \cref{sec:anomaly_detection}, \ie that the attacks are classified as anomalies.
We also verify that our model of normality is valid and does not produce false positives from safe train operation where no attacker is present.

\begin{figure}[htb]
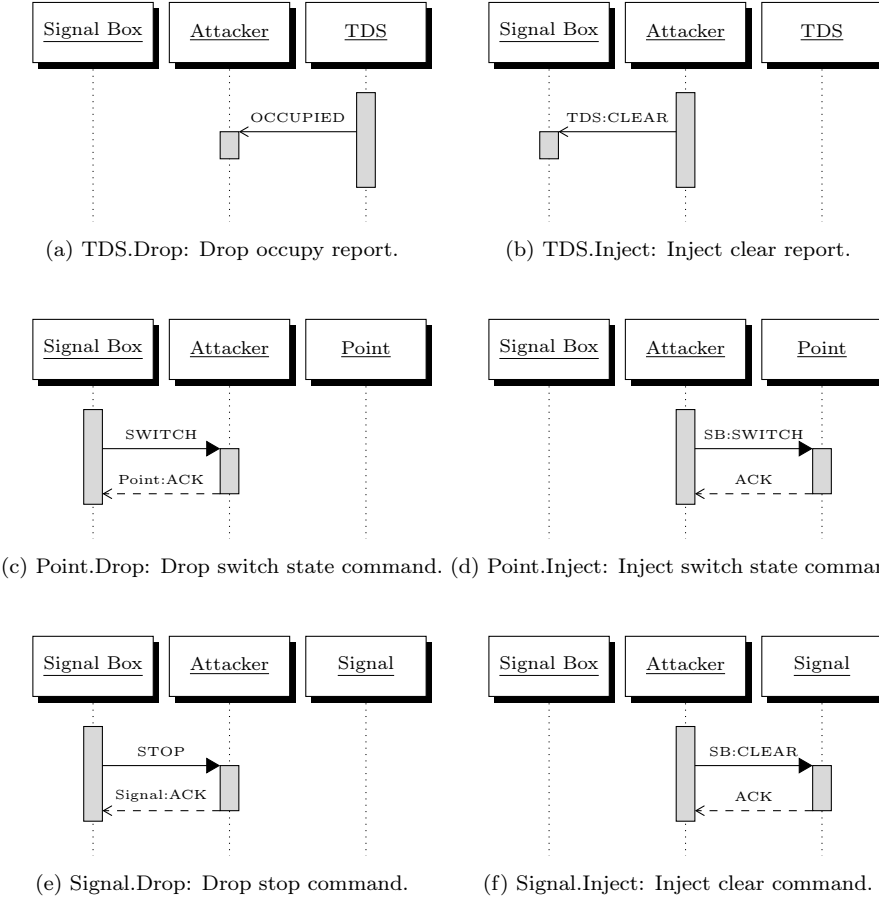

    \centering
    % \tikzset{every picture/.append style={transform shape,scale=.85}}
    \begin{subfigure}[b]{.5\linewidth}
        \centering
        \begin{sequencediagram}
            \tikzset{every node/.append style={font=\scriptsize}}
            \newinst{sb}{Signal Box}
            \newinst{A}{Attacker}
            \newthread{tds}{TDS}

            \begin{messcall}{tds}{\tiny OCCUPIED}{A}
            \end{messcall}

        \end{sequencediagram}
        \caption{TDS.Drop: Drop occupy report.}
        \label{fig:tds.drop}
    \end{subfigure}%
    \begin{subfigure}[b]{.5\linewidth}
        \centering
        \begin{sequencediagram}
            \tikzset{every node/.append style={font=\scriptsize}}
            \newinst{sb}{Signal Box}
            \newthread{A}{Attacker}
            \newinst{tds}{TDS}

            \begin{messcall}{A}{\tiny TDS:CLEAR}{sb}
            \end{messcall}

        \end{sequencediagram}
        \caption{TDS.Inject: Inject clear report.}
        \label{fig:tds.inject}
    \end{subfigure}

    \vspace{2em}

    \begin{subfigure}[b]{.5\linewidth}
        \centering
        \begin{sequencediagram}
            \tikzset{every node/.append style={font=\scriptsize}}
            \newthread{sb}{Signal Box}
            \newinst{A}{Attacker}
            \newinst{p}{Point}

            \begin{call}{sb}{\tiny SWITCH}{A}{\tiny Point:ACK}
            \end{call}

        \end{sequencediagram}
        \caption{Point.Drop: Drop switch state command.}
        \label{fig:point.drop}
    \end{subfigure}%
    \begin{subfigure}[b]{.5\linewidth}
        \centering
        \begin{sequencediagram}
            \tikzset{every node/.append style={font=\scriptsize}}
            \newinst{sb}{Signal Box}
            \newthread{A}{Attacker}
            \newinst{p}{Point}

            \begin{call}{A}{\tiny SB:SWITCH}{p}{\tiny ACK}
            \end{call}

        \end{sequencediagram}
        \caption{Point.Inject: Inject switch state command.}
        \label{fig:point.inject}
    \end{subfigure}

    \vspace{2em}

    \begin{subfigure}[b]{.5\linewidth}
        \centering
        \begin{sequencediagram}
            \tikzset{every node/.append style={font=\scriptsize}}
            \newthread{sb}{Signal Box}
            \newinst{A}{Attacker}
            \newinst{s}{Signal}

            \begin{call}{sb}{\tiny STOP}{A}{\tiny Signal:ACK}
            \end{call}

        \end{sequencediagram}
        \caption{Signal.Drop: Drop stop command.}
        \label{fig:signal.drop}
    \end{subfigure}%
    \begin{subfigure}[b]{.5\linewidth}
        \centering
        \begin{sequencediagram}
            \tikzset{every node/.append style={font=\scriptsize}}
            \newinst{sb}{Signal Box}
            \newthread{A}{Attacker}
            \newinst{s}{Signal}

            \begin{call}{A}{\tiny SB:CLEAR}{s}{\tiny ACK}
            \end{call}

        \end{sequencediagram}
        \caption{Signal.Inject: Inject clear command.}
        \label{fig:signal.inject}
    \end{subfigure}

    \caption{Visualization of attack modes.}
    \label{fig:attack_modes}
\end{figure}

We define attack modes (derived from the attacker's capabilities), building blocks that the attacker uses to desynchronise the signal box's representation of the \ac{fe} states from reality.
To ensure a high level of test coverage, we iterate over the \ac{fe} types and investigate which dropped message and injected message would lead the signalling system into a critical, unsafe state.
The attack modes show \emph{how} the attacker can provoke hazardous situations.
They are then used individually or in combination to conceive the attack scenarios that explain \emph{what} the attacker tries to achieve.
\Cref{fig:attack_modes} shows an overview of all attack modes that create a critical state representation in the signal box.

To reference the attack modes, we name them by the targeted \ac{fe} and the action the attacker performs (inject or drop) separated by a dot.
For TDS.Drop (\cref{fig:tds.drop}), the \enquote{occupied} report is intended to be sent from the \ac{tds} to the signal box but is intercepted and discarded by the attacker.
For TDS.Inject (\cref{fig:tds.inject}), the attacker impersonates a \ac{tds} to counterfeit a cleared track section.
The actual \ac{tds} does not send a message at all.
Both cases make the signal box believe a track section is vacant, while it is in fact occupied.
In turn, to manipulate a command to a point, the attacker needs to forge the respective acknowledgement to not raise suspicion.
Thus, she intercepts and discards the \enquote{switch} command in Point.Drop (\cref{fig:point.drop}) and injects the respective report in the name of the addressed \ac{fe}.
Analogous semantics apply for the visualizations of the remaining attack modes.

From this systematic enumeration of attack modes, we derive the following attack scenarios that we use as test cases for our anomaly detection:

\begin{enumerate}
    \item Switch point under a train (Point.Inject).
    \item Desynchronise point state in the signal box such that a train is diverted into an unsafe route (Point.Drop, Point.Inject).
    \item Allow a train to proceed into a possibly unsafe route:
    \begin{enumerate}
        \item Deviate the track vacancy state in the signal box such that a train is allowed to enter an unsafe route if a signal is set to clear (TDS.Drop, TDS.Inject).
        \item Set a signal protecting an unsafe route to clear or prevent it from setting to stop (Signal.Drop, Signal.Inject).
    \end{enumerate}
\end{enumerate}

In braces, we reference the attack modes used to deploy the scenario.
A route is unsafe if at least a part of it is occupied by another vehicle or reserved for a different, thus conflicting, route.
The attacker can use the defined attack scenarios to achieve her goal of causing a train derailment or collision.

\subsection{Results}
We tested our anomaly detection system on traffic contained in the three datasets containing approximately \num{160 000}, \num{313 000}, and \num{560 000} commands respectively.
To verify the soundness, we ran the system on the datasets without introduced anomalies, where it exhibited no false positives for all three datasets.
Subsequently, we introduced anomalies in the datasets according to the described attack scenarios.
Running the anomaly detection on the modified datasets, we could observe that all anomalies were found, resulting in no false negatives for all three datasets.
By means of simulation, we are able to show that our rule-based anomaly detection provides protection against our attacker model without producing false positives.

The communication overhead of our anomaly detection needs to be put in relation to the timings of the safety process it protects.
We show that our solution does only add small communication overhead to railway signalling.
Setting a real-world point or signal can typically take up to a few seconds.
In fact, setting a route in early generation mechanical signal boxes is estimated to take around \SI{30}{\second} to \SI{120}{\second}, while modern electronic interlocking systems still require about \SI{6}{\second} to \SI{18}{\second}~\cite{pachl2018systemtechnik}.
Hence, the latency introduced by the anomaly detection while setting up a route due to waiting for the query to finish is in the margins of the safety systems.
Additionally, controlling the \ac{fe} is not a time-critical task because the signal box waits for the response of the \acp{fe} in any case, which can take up to a few seconds for a point.

\begin{table}[htb]
    \centering

    \begin{tabular}{r S[table-format=6.2] S[table-format=6.2] S[table-format=6.2]}
        &
        \textbf{Station 1} &
        \textbf{Station 2} &
        \textbf{Station 3} \\
        \midrule
        Field Elements          & 16                  & 29                  & 60                  \\
        Commands                & 161444              & 313177              & 559728              \\
        Algorithm Executions    & 28807               & 51205               & 77734               \\
        Fraction                & 17.84 \si{\percent} & 16.35 \si{\percent} & 13.89 \si{\percent} \\
        Total Messages          & 130812              & 154194              & 465200              \\
        Maximum Messages$^\ast$ & 10                  & 14                  & 18                  \\
        Mean Messages$^\ast$    & 4.54                & 3.01                & 5.98                \\
        Median Messages$^\ast$  & 2.0                 & 2.0                 & 2.0                 \\
    \end{tabular}

    \caption{Communication overhead of the security channel.
        ($^\ast$ Number of messages per query.)}
    \label{tab:communiation_overhead}
\end{table}

\Cref{tab:communiation_overhead} shows detailed figures of our test implementation runs for each tested station.
The first row shows the number of \acp{fe} for each station.
Row \enquote{Commands} contains the number of safety channel commands in our dataset.
\enquote{Algorithm Executions} depicts the number of times our anomaly detection algorithm was started by one of the triggers (see \cref{sec:triggers}), while \enquote{Fraction} is the quotient of the two previous rows to show for how many commands the anomaly detection is executed.
Please note that the share of safety channel messages which trigger the anomaly detection does not exceed \SI{20}{\percent}.
These are commands that switch a point or set a signal to clear, which could set the infrastructure to a critical state.
The remaining messages report \ac{tds} state changes or set a signal to stop, which is not critical according to our model.
Row \enquote{Total Messages} shows the number of security channel messages exchanged.
The remaining three rows indicate the maximum, mean, and median number of security channel messages exchanged per query.
The maximum number of security channel messages is directly determined by the longest route available in the station.
In turn, the longest route is expected to grow disproportionally to the total number of \acp{fe} in the station.
This determines the worst case latency for the security channel.
However, the mean number of security channel messages is much lower than the worst case and grows even slower compared to the number of \acp{fe}.

\begin{mybox}

\paragraph{Really hard-to-defend attacks}

\begin{enumerate}
    \item Train approaches clear signal
    \item Traffic controller sets the signal to stop before train passed (\eg cancelling the route, emergency stop, unsafe conditions)
    \item Attacker drops the stop command and injects a report that the signal shows stop (safety channel compromised)
    \item Train will pass the signal as it still shows clear
\end{enumerate}

Hard to protect as it requires immediate detection of a single missing packet without available context information from the environment (reason for cancelling the route)
Time and target are strictly limited

Another one:

\begin{enumerate}
    \item Train passes a cleared signal
    \item Signal box sends command to set signal to stop
    \item Attacker drops the stop command and injects a report that the signal shows stop (safety channel compromised)
    \item Signal box will believe signal shows stop
    \item The next train approaching the signal will pass it (without permission) and cause a hazard
\end{enumerate}

Not implemented here, but after the adjacent \ac{tds} has been occupied the signal and the \ac{tds} can jointly deduce that the signal should be set to stop and report the missing packet after a timeout.

\end{mybox}

% !TEX root = ../rule-based-anomaly-detection.tex
% !TEX spellcheck = en_GB
% !TEX encoding = UTF-8 Unicode

\section{Discussion}
\label{sec:discussion}

% Realism
Due to the high level of realism of our dataset collected in the signalling lab of TU Darmstadt, our rule-based anomaly detection and its effectiveness can be directly applied to real-world interlocking networks.

\subsection{Freedom of Interference}
We discussed that the fail-safe property of railway signalling is maintained by our anomaly detection because the failure mode is already covered by the safety case (see \cref{sec:handling_alerts}).
Anomaly detection as well as other security measures can be implemented on a \ac{fe}'s controller while maintaining freedom of interference by utilizing a hardware platform as proposed in~\cite{heinrich2019security}.
The hardware platform provides sufficient resources (CPU, memory) to execute our anomaly detection.
A certified hypervisor ensures that the resources of the safety application controlling the \ac{fe} can not be exhausted by other applications, ensuring freedom of interference.

\subsection{Scalability}
Our anomaly detection can be scaled to railway stations of arbitrary size (number of \acp{fe}) because the length of a route typically does not grow in the same order of magnitude as the station's size.
This can be seen in \cref{tab:communiation_overhead} by comparing the number of \acp{fe} (station size) to the maximum and mean number of messages per query (route length).
Additionally, the queries of multiple routes are executed in parallel and do not consume significant network bandwidth because they require only a few bytes (state information plus, \eg \ac{mac} for authenticity).

\subsection{Limitations}
% Limitations: overlap, approach signals
There are several special cases of train operation that are not yet covered by the ruleset presented in this article.
After the exit signal of a route, the consecutive section, the so-called overlap, needs to be vacant as well.
The overlap is reserved in case the train is unable to stop at the signal and slips past it.
This is not reflected with our anomaly detection.
Furthermore, approach signals forecast the aspect of a main signal to accommodate for the long breaking distances of trains.
A check needs to be performed that the approach signal always shows the signal aspect respective to the aspect of the following main signal.
This case needs to consider that an approach signal can be assigned to multiple main signals and vice versa.
We do not consider overlap, approach signals, and shunting signals in this paper for simplicity, but plan to handle them and more special cases in future work.

\subsection{Merits}
Our anomaly detection makes railway signalling resilient to the two attacker types described in \cref{sec:attacker_model}: a \dy attacker in the communication network issuing licit but mistimed command an control messages as well as an attacker who compromises the signal box which holds and evaluates the safety logic.
In particular, the \dy attacker can control all safety channels without being able to implement a successful attack by dropping or injecting control messages.
Any attempt to set the infrastructure to an unsafe state is prohibited by the rules of the anomaly detection (security channel) as we evaluated with our attack scenarios in \cref{sec:evaluation}.

Additionally, the attacker can gain control over the signal box, which gives her full control over the supervised area and the power to cause severe damage.
With the proposed anomaly detection system in place, the actions of the signal box are validated against the ruleset independently and hazardous commands rejected on each \ac{fe}, thus fully mitigating the possibilities of a rogue signal box to create critical states.

Changes to the safety functionality of the railway system are required to undergo a lengthy admission process that conflicts with typical security update cycles.
This \enquote{update problem}, as identified in several other publications~\cite{valdivia2018cybersecurity,schlehuber2017security,schlehuber2017challenges}, is avoided by our anomaly detection.
Our rules are derived once and are static such that updates are not required.
The anomaly detection is general enough to be applied to arbitrary railway stations independent from their concrete topology, which we showed by applying it to three different topologies in our evaluation.
Hence, our rule-based anomaly detection can be re-used by configuring the security channels once for any new station.

\begin{mybox}

\subsection{Limitations}

\begin{itemize}
    \item The \enquote{Really hard-to-defend attacks} could be discussed as limitation
\end{itemize}

\subsubsection{Difference to Arwed}

\begin{itemize}
    \item No approach signal implemented here
    \item Replaced response collection time with security channels
    \item Unlock routes if signal is set to stop
    \item Implicitly store open line and terminal status of track section in successors (None reference)
    \item Store pred/succ neighbours (replaces lookup in find\_next)
\end{itemize}

\subsubsection{Dataset}
\label{sec:dataset}

Some records that cause command rejections seem to be swapped.
They appear in the same second, but a point is cleared after the command to switch is issued.
In reality this could be in the correct order (cleared, then switch).
What we see could simply be a race condition in the EBD.

Reasons for rejected routes:
\begin{itemize}
    \item Train has gone a different way than point states suggest (unused element remains locked)
    \item Train disappears (occupation of track section is missing) (unused element remains locked)
    \item Signal gets set to clear multiple times (elements are already locked). Check whether aspect actually changes or goes from clear to clear.
    \item Point is switched and then cleared immediately after in the dataset. Commands could simply be swapped.
\end{itemize}

\end{mybox}

% !TEX root = ../rule-based-anomaly-detection.tex
% !TEX spellcheck = en_GB
% !TEX encoding = UTF-8 Unicode

\section{Conclusion}
\label{sec:conclusion}

We described a distributed rule-based anomaly detection system dedicated to the \acl{ci} of railway transportation.
As one of several security measures in a \did concept for a safety-critical system, the anomaly detection mitigates the threats of two attacker types: an attacker who compromised a railway station's signal box that contains the safety logic and a \dy attacker who gained access to the signalling network.
Both attacker types aim to cause a train collision or derailment.
Our work shows that anomaly detection is an effective defence strategy which can be applied to railway stations of arbitrary topology and size.
It inspects control commands at the receiving \ac{fe} and verifies their harmlessness by applying knowledge about the current state of neighbouring \acp{fe} along the route the train is supposed to travel.
If a hazardous command is detected, the anomaly detection system prevents the execution of the command with minimal interference with the safety functionality and raises an alert about the incident.
Small overhead is produced by allowing \acp{fe} to communicate with each other that are immediate neighbours in the railway topology.

\begin{mybox}

neighbours only
query following the route
complexity reduction with signal mapping

\begin{itemize}
    \item Security channel helps with the long lifetime of signalling equipment and covers legacy systems
    \item Redundancy is important in safety-critical systems
    \item it is impossible to monitor the entire infrastructure
\end{itemize}

\end{mybox}

% ---------- Acknowledgements ----------
\section*{Acknowledgements}

The work presented in this paper has been partly funded by the German Federal Ministry of Education and Research (BMBF) under the project \enquote{HASELNUSS: Hardwarebasierte Sicherheitsplattform für Eisenbahn-Leit- und Sicherungstechnik} (ID 16KIS0597K).

% ---------- Bibliography ----------
\section*{\refname}
\bibliographystyle{elsarticle-num}
\bibliography{bibliography}

\ifcomments
% !TEX root = ../rule-based-anomaly-detection.tex
% !TEX spellcheck = en_GB
% !TEX encoding = UTF-8 Unicode
\newpage
\section*{Unused Material}

\begin{figure}[htb]
    \centering
    \includegraphics[width=\textwidth]{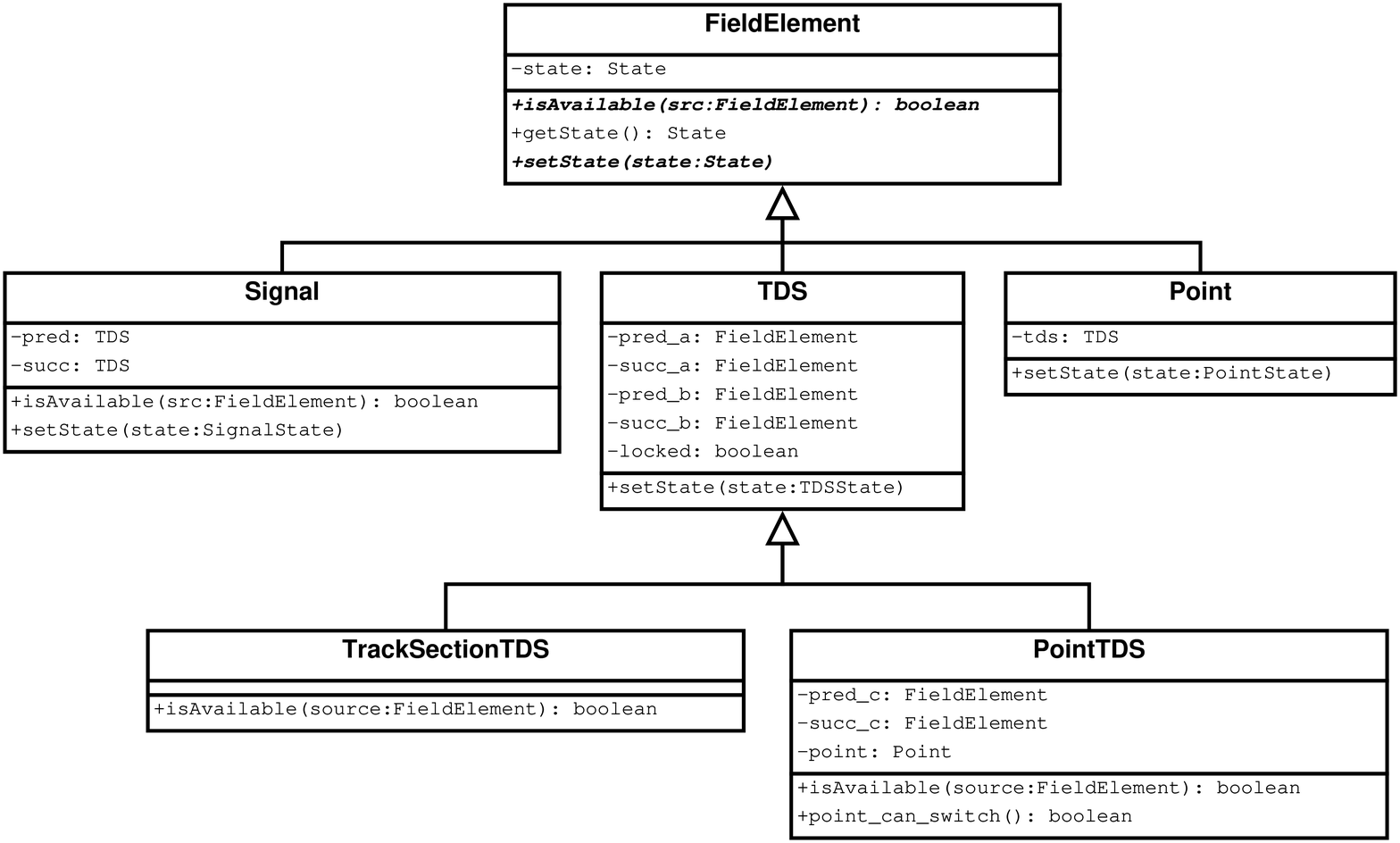}
    \caption{Class diagram.}
    \label{fig:class_diagram}
\end{figure}

A topology that can appear but might not be reflected in Arwed's thesis.
An approach signal is associated with more than one main signal because there is a point between them.
The approach signal needs a method to find the correct main signal (\cref{fig:approach_signal_edge_case}).
Arwed's proposed methodology assumes a 1:1 relation and uses a direct reference from approach signal to main signal.
For main signals that can show approach aspects, Arwed assumes that a respective route has been set up before.
The main signal stores a reference to the signal at the end of the route which is naturally the signal to show the approach aspect for.
The reference is determined during the availability query process when the route is set.
The query ends at the respective signal.

\begin{figure}[htb]
    \centering
    \includegraphics{gfx/Approach_Signal_Edge_Case.pdf}
    \caption{Edge case with approach signals.}
    \label{fig:approach_signal_edge_case}
\end{figure}

\listoftables

\listoffigures
\fi

% ---------- Build Information ----------
\FloatBarrier
\vfill
{\scriptsize
\DTMNow \hfill
}

\end{document}